\documentclass[12pt,preprint]{aastex}
\slugcomment{Accepted for publication in the ApJ: July 15, 2005}

\newcommand       \mum        	 {$\mu$m}

\newcommand	     \zmin      	{$z_{min}$}
\newcommand      \zmax      	{$z_{max}$}

\newcommand      \gray       {$\gamma$-ray}
\begin{document}

\title{IS THERE AN IMPRINT OF PRIMORDIAL STARS IN THE TeV $\gamma$-RAY SPECTRUM OF BLAZARS?}

\author{Eli Dwek\altaffilmark{1}, Frank Krennrich\altaffilmark{2}, and
Richard G. Arendt\altaffilmark{3}}
\altaffiltext{1}{Observational Cosmology Laboratory, Code 665, NASA Goddard Space Flight Center,
Greenbelt, MD 20771,  e-mail: eli.dwek@nasa.gov}
\altaffiltext{2}{Department of Physics and Astronomy, Iowa State University, Ames, IA 50011}
\altaffiltext{3}{Science Systems and Applications, Inc., Observational Cosmology Laboratory, Code 665, NASA Goddard Space Flight Center, Greenbelt, MD 20771}

\begin{abstract}
The 1 - 5 \mum\ diffuse sky emission from which local foreground emission from the solar system and the Galaxy have been subtracted exceeds the brightness that can be attributed to normal star forming galaxies. The nature of this excess near-infrared background light (NIRBL) is still controversial. On one hand, it has been interpreted as a distinct spectral feature created by the redshifted emission from primordial (Population~III) stars that have formed at redshifts $\gtrsim$ 8. On the other hand, the NIRBL spectrum is almost identical to that of the zodiacal cloud, raising the possibility that it is of local origin. Blazars can, in principle, offer a simple test for the nature and origin of the NIRBL. Very high energy \gray\ photons emitted by these objects are attenuated on route to earth by $\gamma$-$\gamma$ interactions with the extragalactic background light (EBL). Assuming that the NIRBL is of extragalactic origin, its distinct spectral feature should give rise to a corresponding absorption feature in the observed $\gamma$-ray spectra of these sources.  This paper examines whether the extragalactic nature of the NIRBL can be determined from the analysis of the \gray\ spectra of blazars. We calculate the \gray\ opacity to two blazars H1426+428 and PKS2155-304 located at redshifts of $z \approx$ 0.13, for several EBL scenarios with and without the alleged spectral signature of  the Population~III stars. When the NIRBL is assumed to be extragalactic, the \gray\ opacity rises above unity with a distinct jump at around 300~GeV. Applied to H1426+428, the NIRBL does not produce any distinct spectral features in the absorption-corrected spectrum of this blazar (since its observed spectrum extends down to only $\sim$ 300 GeV), but causes an unlikely rise in its spectral behavior with energy.  A more interesting result is obtained for the blazar PKS2155-304 which has a more accurately determined \gray\ spectrum, well approximated by a power law, that also extends to a lower energy of $\sim$ 200~GeV. Because of its broader energy coverage, the increased TeV opacity (when the NIRBL is assumed to be extragalactic), gives rise to a parabolic-shaped absorption-corrected spectrum peaking at about 1-2~TeV with a possible inflection around $\sim$ 300~GeV. Corrected only for absorption by the EBL produced by galaxies (i.e. without the NIRBL), the spectrum of PKS2155-304 is simply a power law. These results illustrate that if a blazar is located at a sufficiently large redshift, the NIRBL can produce a distinct spectral feature in its \gray\ spectrum. Without any a priori knowledge of its intrinsic spectrum, there is no reason to prefer one absorption-corrected spectrum over the other. However, it will require a very fine tuning of the EBL spectrum to convert an intrinsic blazar spectrum with an inflection at 300~GeV to an observed  power law spectrum. Furthermore, we show that  if the NIRBL is extragalactic, it may be very difficult to reproduce the absorption corrected spectrum of PKS2155-304 with a synchrotron self Compton (SSC) model that is also consistent with the X-ray and EGRET \gray\ data. All these arguments cast serious doubts on the possibility that the NIRBL represents the spectral imprint of the first generation of stellar objects on  the EBL. 
\end{abstract}

\section{INTRODUCTION}
Observations of TeV blazars offer the exciting possibility of detecting an absorption signature in their spectra that can be directly attributed to the radiative contribution of the first generation of stars to the extragalactic background light (EBL).  The radiative output  of these stars leaves a distinct spectral signature at near-IR wavelengths (Santos, Bromm, \& Kamiensky 2003), which may have been detected by the Diffuse Infrared Background Experiment (DIRBE) instrument on board the {\it Cosmic Background Explorer} ({\it COBE}) satellite (Hauser et al. 1998), and the Near Infrared Spectrometer (NIRS) instrument on board the {\it Infrared Telescope in Space} ({\it IRTS}) (Matsumoto et al. 2000, 2004) satellites. The diffuse sky emission from which foreground emissions from the zodiacal dust cloud and galactic starlight has been subtracted show an excess of emission above that attributed to normal star forming galaxies (Matsumoto et al. 2004; Totani et al. 2001, Salvattera \& Ferrara 2004; Dwek \& Arendt 2005). The nature of this emission is still unresolved. Dwek \& Arendt (2005) have suggested that this excess near-infrared emission could be caused by an unexpectedly large isotropic component that was not included in the model used to subtract the zodiacal light from the DIRBE and NIRS data. However, the detection of spatial fluctuations in this residual component in excess of that expected from the zodiacal cloud, the Milky Way, and normal galaxies, suggests that it may be produced by a population of primordial stars (Population~III; Pop~III stars). We hereafter refer to the excess 1 - 5 \mum\ near-infrared background light from which the zodiacal light, Galactic stars, and the contribution from star-forming galaxies have been subtracted as the NIRBL. Assuming the NIRBL to be extragalactic, Salvaterra \& Ferrara (2004) and Dwek \& Arendt (2005) have fitted its spectrum with the cumulative radiative output from Pop~III stars, that were continuously formed between redshifts of \zmax $\approx$ 15--30 and \zmin $\approx$ 7--9. The NIRBL has a distinct spectral signature, characterized by a sharp rise between 0.8 and 1.25 \mum, followed by a $\lambda^{-2}$ decline to $\sim$ 5 \mum. If the NIRBL is indeed extragalactic, then it will leave its imprint on the \gray\ opacity toward distant blazars, and affect the shape of their observed spectra. 

Mapelli, Salvaterra, \& Ferrara (2004) have suggested that the observed "S"-shaped TeV spectrum of the blazar H1426+428 represents the absorption signature of the intrinsic blazar spectrum, which they a priori assumed to be a power law. However,  Dwek \& Krennrich (2005) showed that blazars cannot, in  general, be used to discriminate between many possible realizations of the EBL, since their intrinsic \gray\ spectra are a priori unknown. Nevertheless, it is of great interest to examine in more detail if the observed \gray\ spectra of blazars can indeed be used to prove or disprove the extragalactic nature of then NIRBL. The zodi- and Galaxy-subtracted 1 - 5 \mum\ DIRBE and NIRS diffuse sky measurements were previously used to create different realizations of the EBL which were used to explore the physical characteristic of the absorption-corrected spectra of blazars (Dwek \& Krennrich 2005, Costamante et al. 2004, and Aharonian et al. 2002). However, none of these studies attributed the $\sim$ 1 - 5 \mum\ EBL intensities to the distinct emission from Pop~III stars. Consequently, the EBL realizations fitted through these data points did not have the unique NIRBL spectral signature depicted in Figure 2. 

We first summarize the current limits and detections of the EBL in the 0.1 to 10 \mum\ wavelength region, and present the model fits of Dwek \& Arendt (2005) to the NIRBL (\S2) In \S3 we derive the absorption-corrected \gray\ spectra of four blazars: H1426+428 (1ES1268+428), located at reshift $z$ = 0.129; PKS2155-304, located at $z$ = 0.117; and the two relatively nearby blazars Mrk~421 and Mrk~501, located at $z$ = 0.030 and 0.033, respectively. The spectra are corrected for absorption using several different realizations of the EBL: (1) an EBL consisting only of the emission from normal star forming galaxies (hereafter GEBL), i.e. we assume that in this case the NIRBL is due to a foreground emission component; and (2) an EBL that consists of the GEBL and two different model fits to the NIRBL (Dwek \& Arendt 2005). The astrophysical implications of our results are discussed in \S4.

In all our calculations we adopt a cosmological model with parameters determined from the analysis of the {\it Wilkinson Microwave Anisotropy Probe} ({\it WMAP}, Bennett et al. 2003): a dark energy density of $\Omega_{\Lambda}$ = 0.73,  and total and baryonic matter densities of $\Omega_m$ = 0.27 and $\Omega_b$ = 0.044, respectively. The densities are normalized to the critical density using a Hubble constant of $H_0$ = 100$h$~km~s$^{-1}$ with $h$ = 0.70.

 \section{CURRENT LIMITS AND DETECTIONS OF THE EXTRAGALACTIC BACKGROUND LIGHT BETWEEN 0.1 AND 10 \mum.}
 Figure 1 depicts the current limits and detections of the EBL in the 0.1 to 10 \mum\ wavelength region. The data include: (1) ground- and space-based measurements of integrated galaxy light (Madau \& Pozzetti 2000, open triangles; Bernstein, Freedman, \& Madore 2002, open diamonds; Fazio et al. 2004, open squares); (2) direct measurements based on data obtained by the {\it COBE}/DIRBE (Hauser et al. 1998; Dwek \& Arendt 1998; Arendt \& Dwek 2003, filled diamonds; Cambr\'esy et al. 2001, filled triangles; Wright \& Reese 2000; Gorjian et al. 2000; and Wright 2001, filled squares), and the {\it IRTS}/NIRS (Matsumoto et al. 2000, 2004, filled circles) instruments; and (3) extrapolated galaxy number counts (Totani et al. 2001, open circles). 
 The thick solid line in the figure is a polynomial fit to the EBL formed by star forming galaxies, as defined by the Totani et al. (2001) model.  
The polynomial fit includes the mid- and far-IR limits and measurements by Infrared Array Camera (IRAC) intsruments on board the {\it Spitzer} satellite (Fazio et al. 2004, open squares); Metcalfe et al. (2003, 15 \mum), Papovich et al. (2004, 24 \mum), Lagache et al. (2000, 100 \mum), and Hauser et al. (1998, 140, 240 \mum). We refer to this EBL model hereafter as the GEBL.

  \begin{figure}
  \epsscale{0.5}
 \plotone{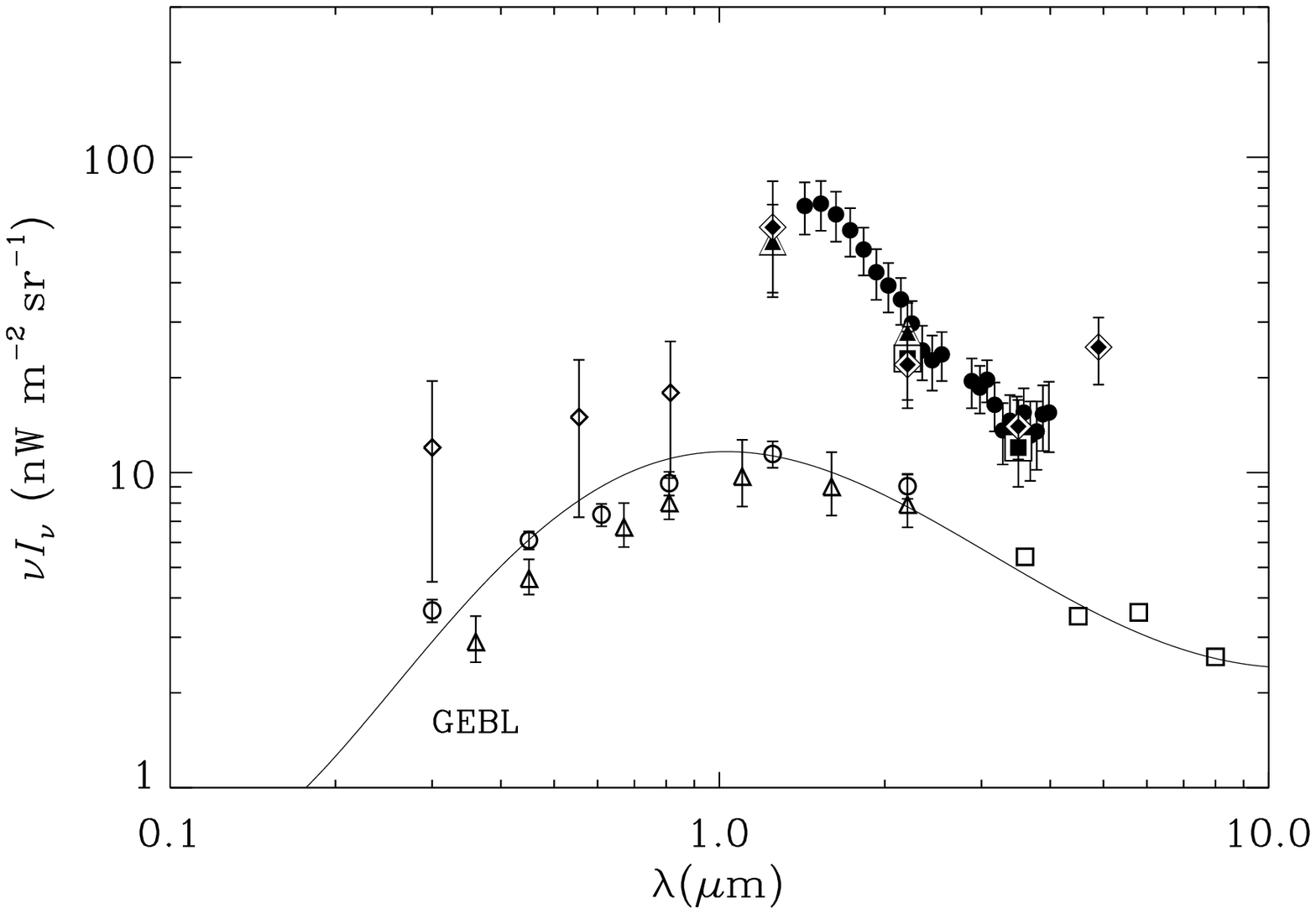}
 \caption{Limits and detections of the EBL. The thick solid line is the nominal EBL spectrum attributed to "normal" star forming galaxies. References to the observations are in the text.}
  \epsscale{1.0}
 \end{figure}
 
 Figure 2 shows the two separate fits to the NIRBL corresponding to the two Pop~III star formation scenarios considered by Dwek \& Arendt (2005). In the first scenario, the Pop~III stars form over the redshift interval \{\zmin, \zmax\}=\{7, 15\}, and in the second they form over the  \{\zmin, \zmax\}=\{9, 30\} redshift interval.  The DIRBE and NIRS observations do not provide tight constraints on the value of \zmin\ since the short wavelength cutoff of the NIRBL is only constrained to lie between 0.8 and 1.25 \mum.  Salvaterra \& Ferrara (2003) attributed great statistical significance to the apparent drop in the 1.25 \mum\ DIRBE derived flux compared to the NIRS spectrophotometric data, deriving a value of \zmin\ = 8.8. We do not regard the drop in the 1.25 \mum\ flux statistically significant because of the large error bar on the 1.25~\mum\  data point and the presence of systematic uncertainties between the two DIRBE and NIRS instruments. 
 Designating the NIRBL corresponding to the two Pop~III formation scenarios by NIRBL7 and NIRBL9, respectively,  we can define two more EBL realizations formed by the addition of each of these NIRBL spectra to the GEBL.  

  \begin{figure}
 \plottwo{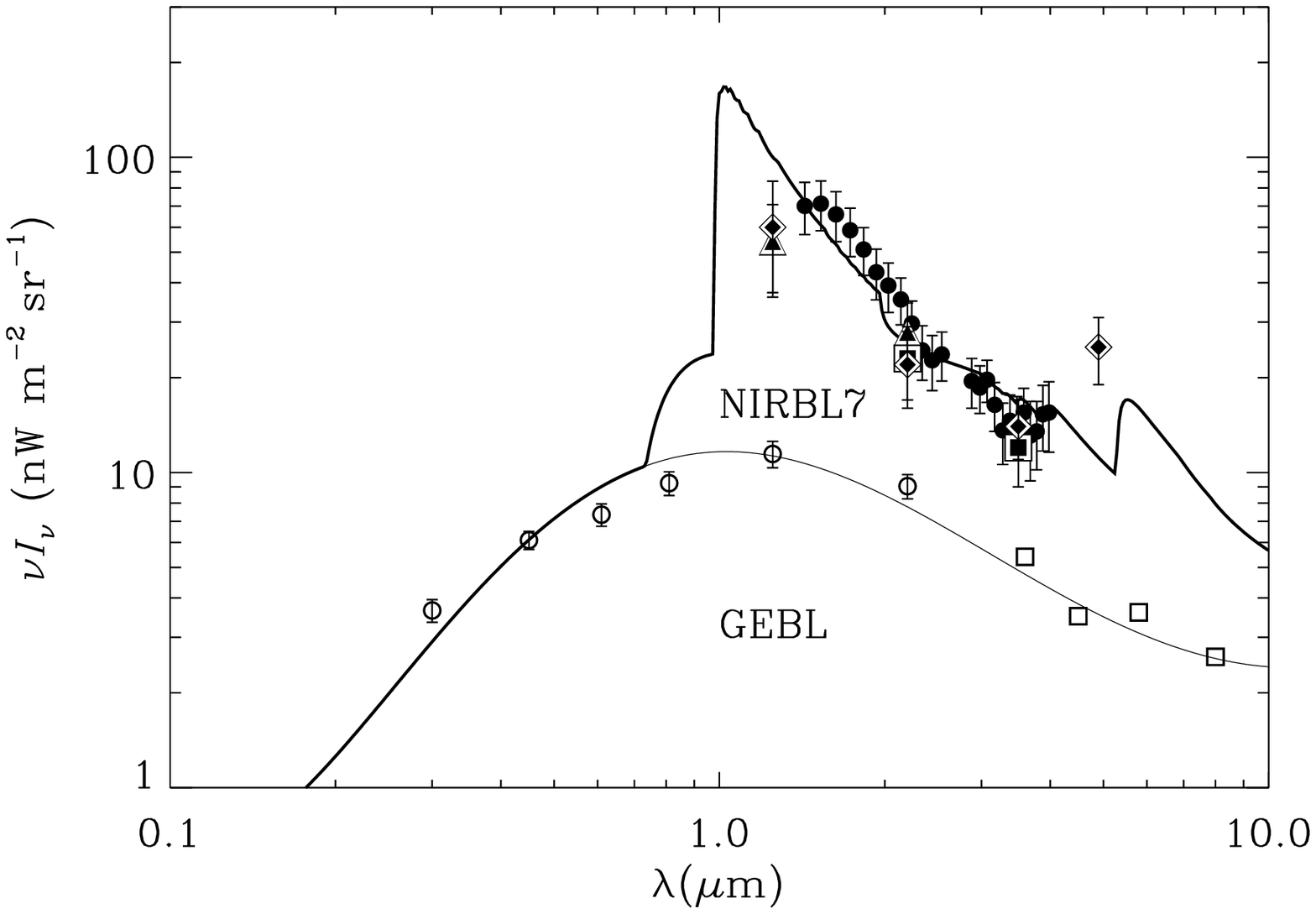}{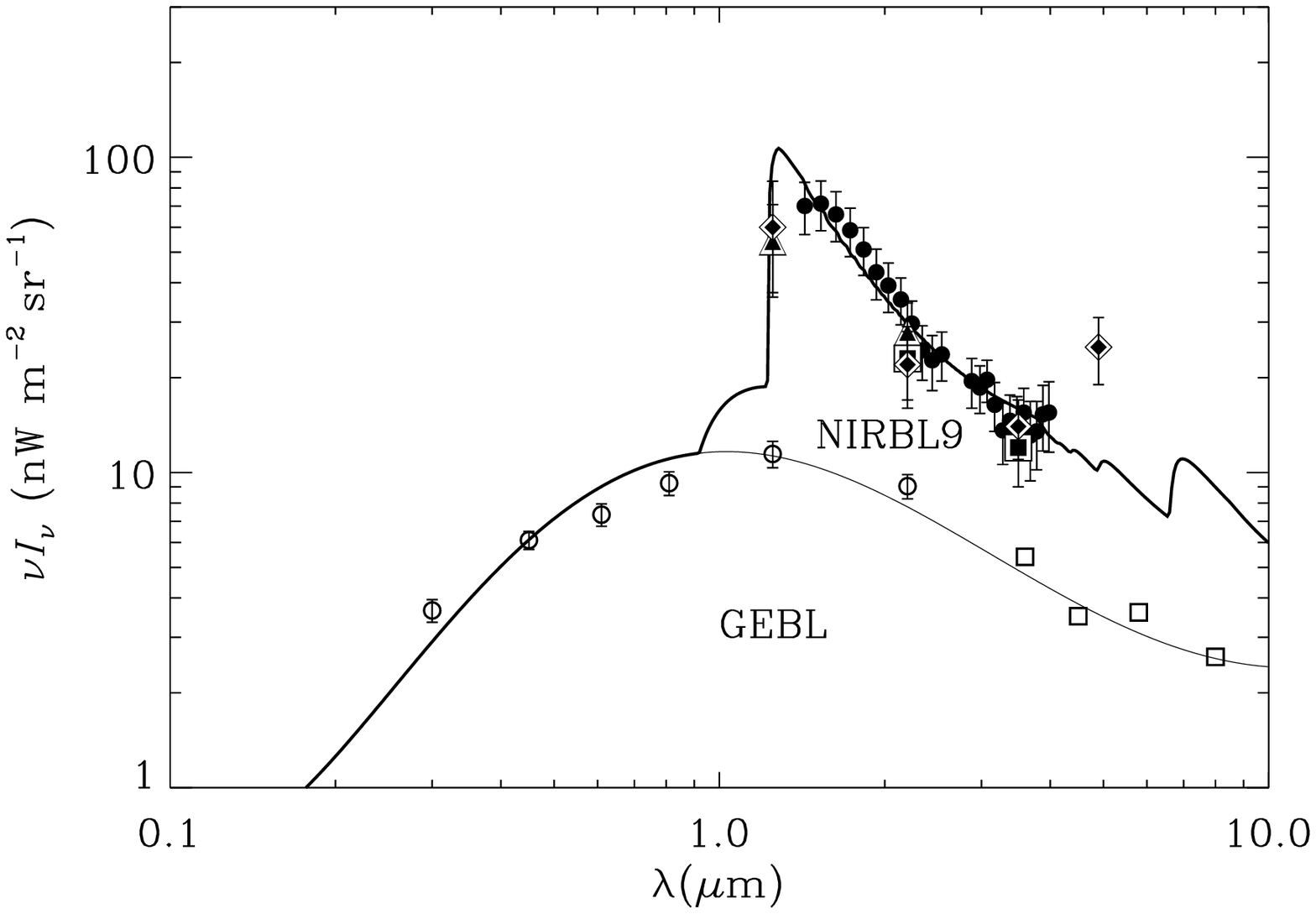}
  \caption{The imprint of Pop III stars on the EBL for different formation scenarios characterized by \{$z_{min}$, $z_{max}$\} = \{7, 15\} (left panel) and  \{$z_{min}$, $z_{max}$\} = \{9, 30\} (right panel); see Dwek \& Arendt (2005) for more details. The NIRBL is the {\it excess} near-infrared background light over that defined by the GEBL, and is given by the areas marked NIRBL7 and NIRBL9, corresponding to the Pop~III star formation scenarios.}
 \end{figure}


 \section{THE $\gamma$-RAY OPACITY FOR DIFFERENT EBL REALIZATIONS}

Figures 3  depicts the GEBL+NIRBL7 realization described above (left panel), and the \gray\ opacity to a source at redshift  $z$ = 0.122, which is the average redshift for the two distant blazars H1426+428 and PKS2155-304 (right panel).  For more nearby blazars, the optical depth scales simply linearly with redshift. Formulae for calculating the optical depth are given, for example, by Dwek \& Krennrich (2005).
The shaded curves in the right panel represent the contribution of the different wavelength regions to the total \gray\ opacity. Figure 4 shows the same quantities for the GEBL+NIRBL9 realization of the EBL.

  \begin{figure}
 \plottwo{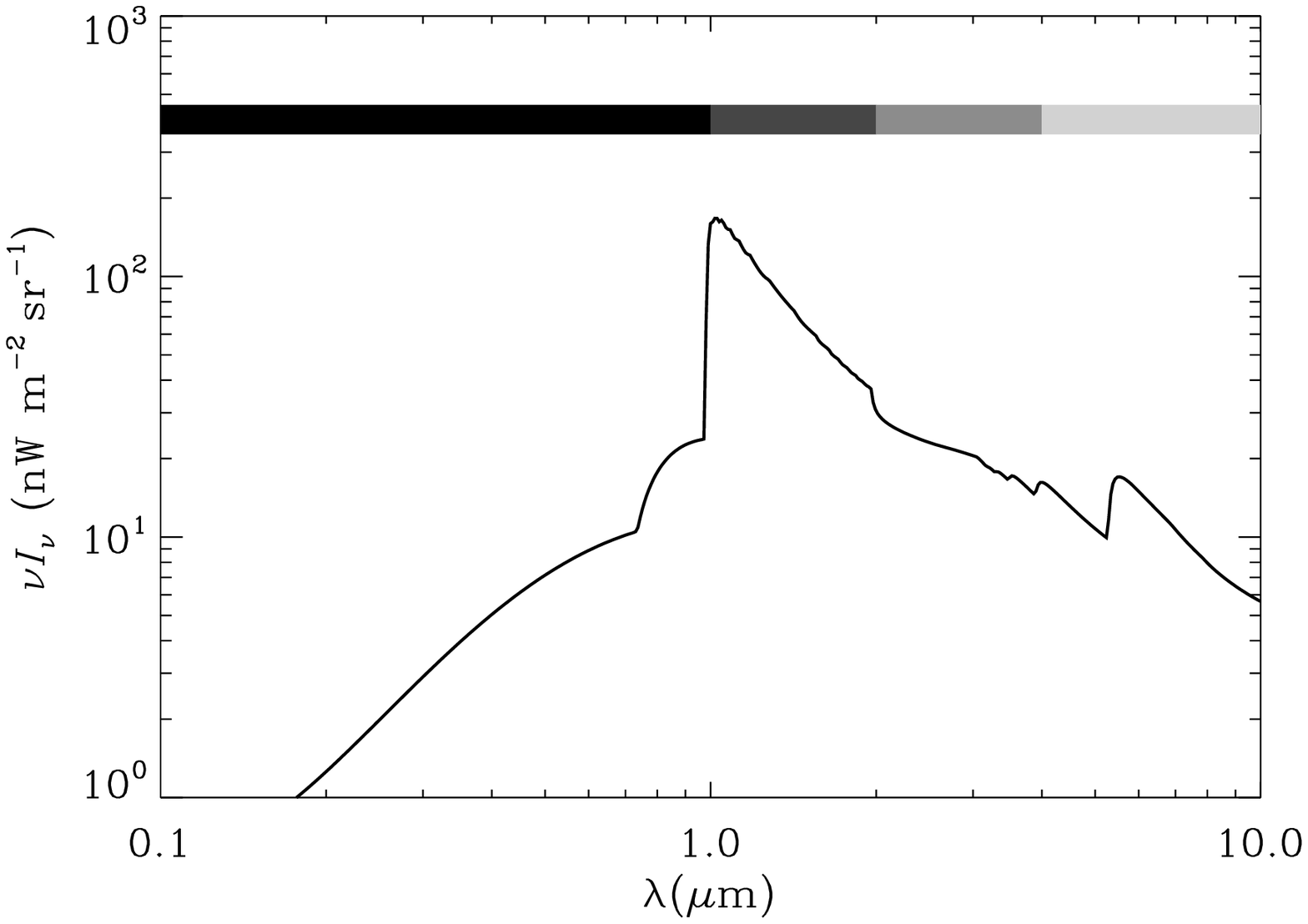}{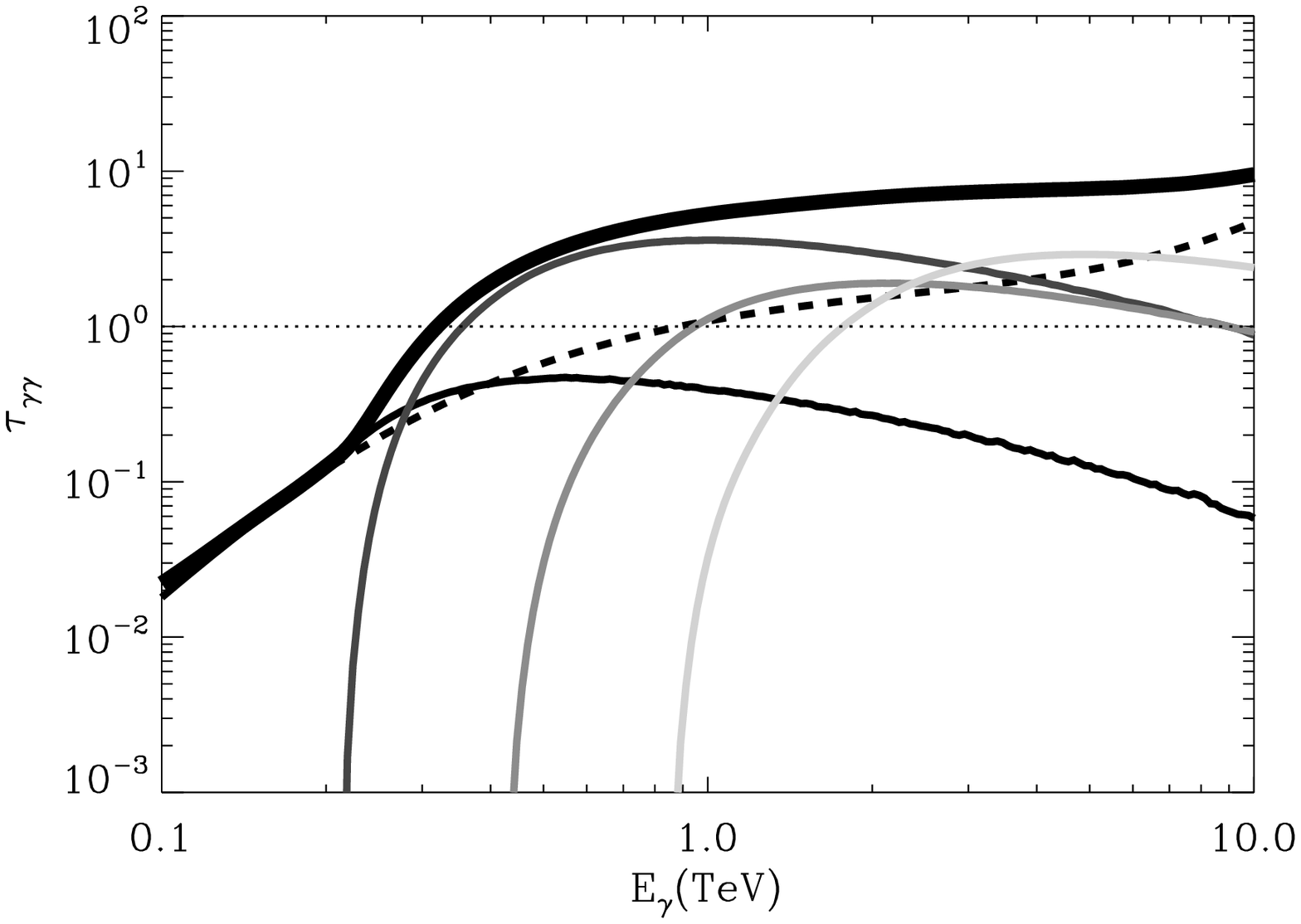}
  \caption{{\it Left panel}: The EBL background consisting of the GEBL and NIRBL7. {\it Right panel}: The \gray\ opacity to a source at redshift $z$ = 0.122 due to $\gamma$-$\gamma$ interactions with the EBL. The shaded curves in the figure represent the contribution of the different wavelength regions (shown by the same shaded area in the left panel) to the total \gray\ opacity. The dashed line represents the \gray\ opacity caused by the GEBL alone (see Figs 1 and 2).}
 \end{figure}

  \begin{figure}
 \plottwo{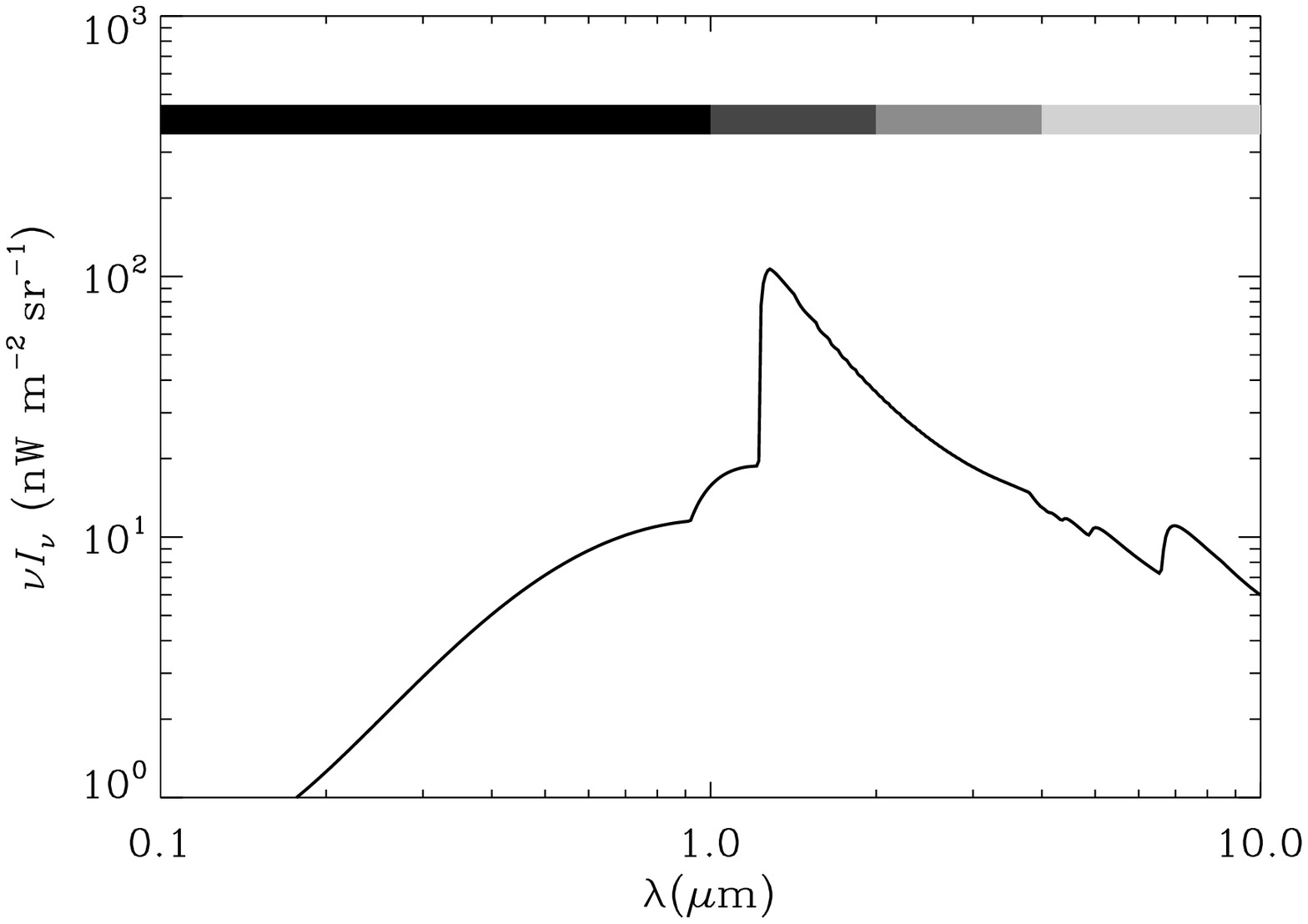}{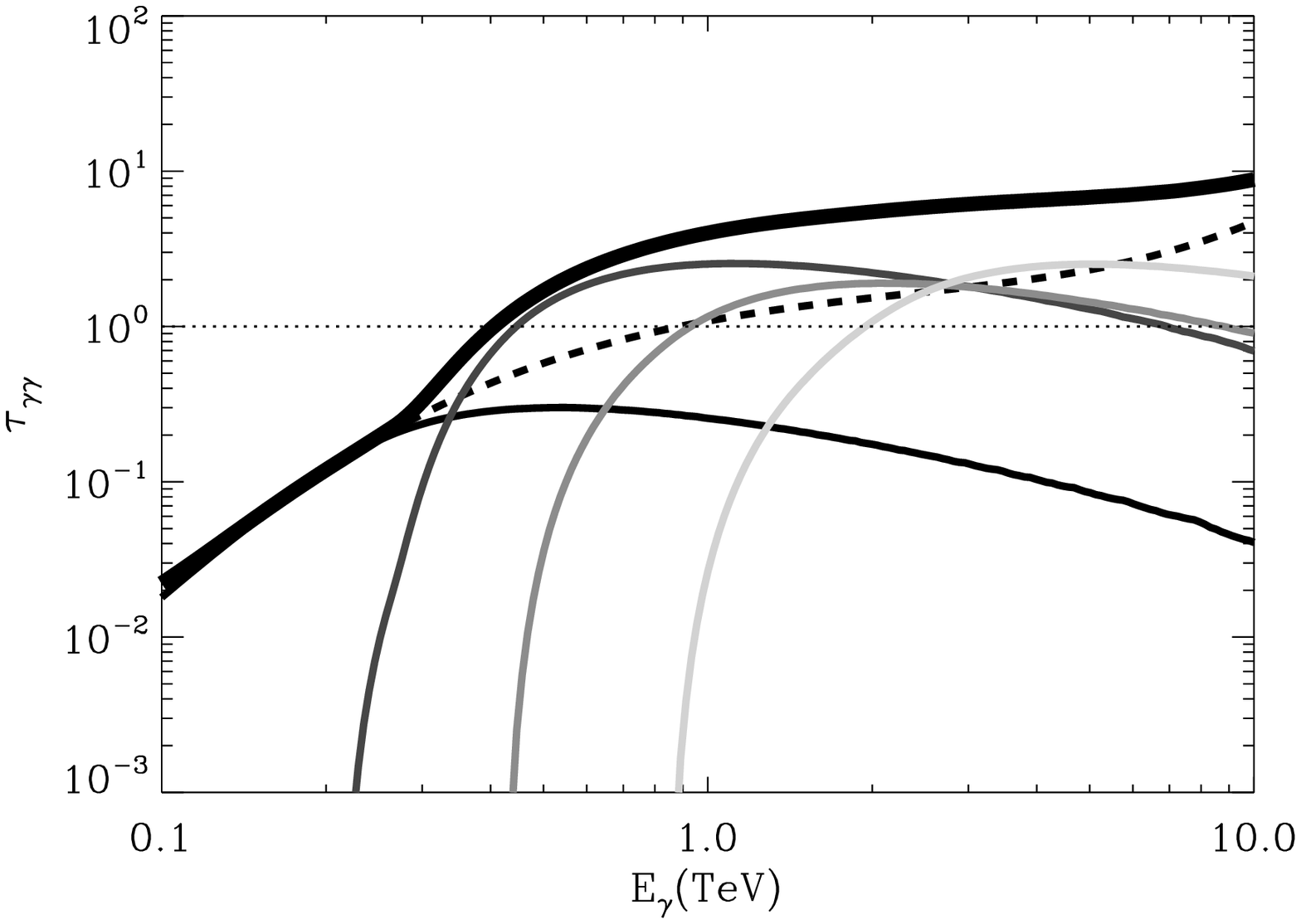}
  \caption{Same as Figure 3 for the GEBL and NIRBL9 realization of the EBL.}
 \end{figure}

The figures show that the effect of the NIRBL is to produce a sudden rise in the \gray\ opacity to these two sources at around 300~GeV, compared to the more gradual rise in the opacity caused by the GEBL alone (dashed line in the figures). This discontinuous rise is caused by the relatively small 200~GeV opacity produced predominantly by the GEBL, followed by an increase in the opacity at higher energies caused by the additional contribution of the NIRBL to the \gray\ opacity. 
The figures also demonstrate that at $z \sim$ 0.12, the sources are sufficiently far away for the NIRBL to cause the opacity to rise above unity at an energy of  $\sim$ 300~GeV, instead of $\sim$ 1~TeV if the EBL consisted of only the GEBL. This effect will produce distinctly different absorption-corrected spectra for the GEBL and GEBL+NIRBL realizations of the EBL.
Another important thing to note is that the sharp discontinuity in the spectral shape of the NIRBL is not translated into a comparable sharp discontinuity in the \gray\ opacity. This is a direct result of the rather broad \gray\ cross section which will tend to smooth out any sharp features in the EBL. This effect is illustrated in Figure 5 where we compare the observed spectrum of H1426+428 to that expected from a blazar with an $dN/dE \propto E^{-1}$ intrinsic spectrum being absorbed by the GEBL+NIRBL7. The figure shows the effect of the smoothing of the NIRBL when it is convolved with the cross section for the $\gamma$-$\gamma$ reaction. Mapelli et al. (2004) have attributed great significance to the peculiar shape of the observed \gray\ spectrum taking it as evidence for the effect of absorption by the NIRBL. This seems to be supported by the apparently good fit of the (GEBL+NIRBL)-corrected spectrum to the data. Figure 5 also depicts the effect of EBL absorption when the intrinsic blazar spectrum is given by a  $dN/dE \propto E^{-7/3}$ power law, and the EBL consists of only the GEBL. The resulting observed spectrum is shown by a dashed line in the figure, and produces a similarly good fit to the observed blazar spectrum as the previous case. We therefore conclude that, although the shape of the observed spectrum is definitely sensitive to the detailed spectrum of the EBL in the 1 to 15~\mum\ wavelength regime (Aharonian et al. 2002, 2003), it does not provide any evidence for the extragalactic nature of the NIRBL. 

  \begin{figure}
   \epsscale{0.5}
 \plotone{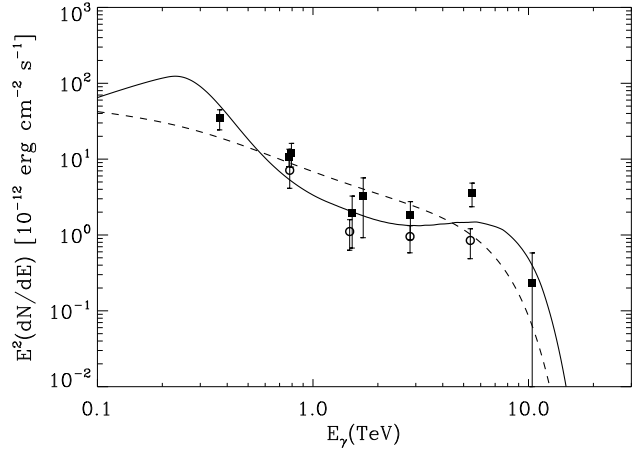}
 \caption{The observed spectrum of H1426+428 is compared to that expected from a blazar with an intrinsic spectrum: $dN/dE \propto E^{-1}$ corrected for absorption by a GEBL+NIRBL7 realization of the EBL (solid line). The dashed line represents the observed spectrum corrected for absorption by the GEBL, assuming the intrinsic spectrum is given by $dN/dE \propto E^{-7/3}$. The figure shows that both EBL realizations give equally good fit to the data presented by Petry et al (2002; filled squares) and data shown by Aharonian et al (2003; open circles).}
  \epsscale{1.0}
 \end{figure}

\subsection{H1426+428}
The blazar H1426+428 was detected at TeV energies by the VERITAS (Petry et al. 2002), and HEGRA (Aharonian et al. 2003) collaborations. 
Figure 6 depicts the observed \gray\ spectrum of H1426+428 corrected for the GEBL and the GEBL+NIRBL7 and GEBL+NIRBL9 realizations of the EBL (left panel). Without any a priori knowledge of its intrinsic spectrum, all absorption-corrected spectra could equally well be the intrinsic H1426+428 spectrum. Consequently, there is no possible way to use the currently observed blazar spectrum to prove the extragalactic nature of the NIRBL, a point already illustrated in the previous section and in Figure 5. 

  \begin{figure}
 \plottwo{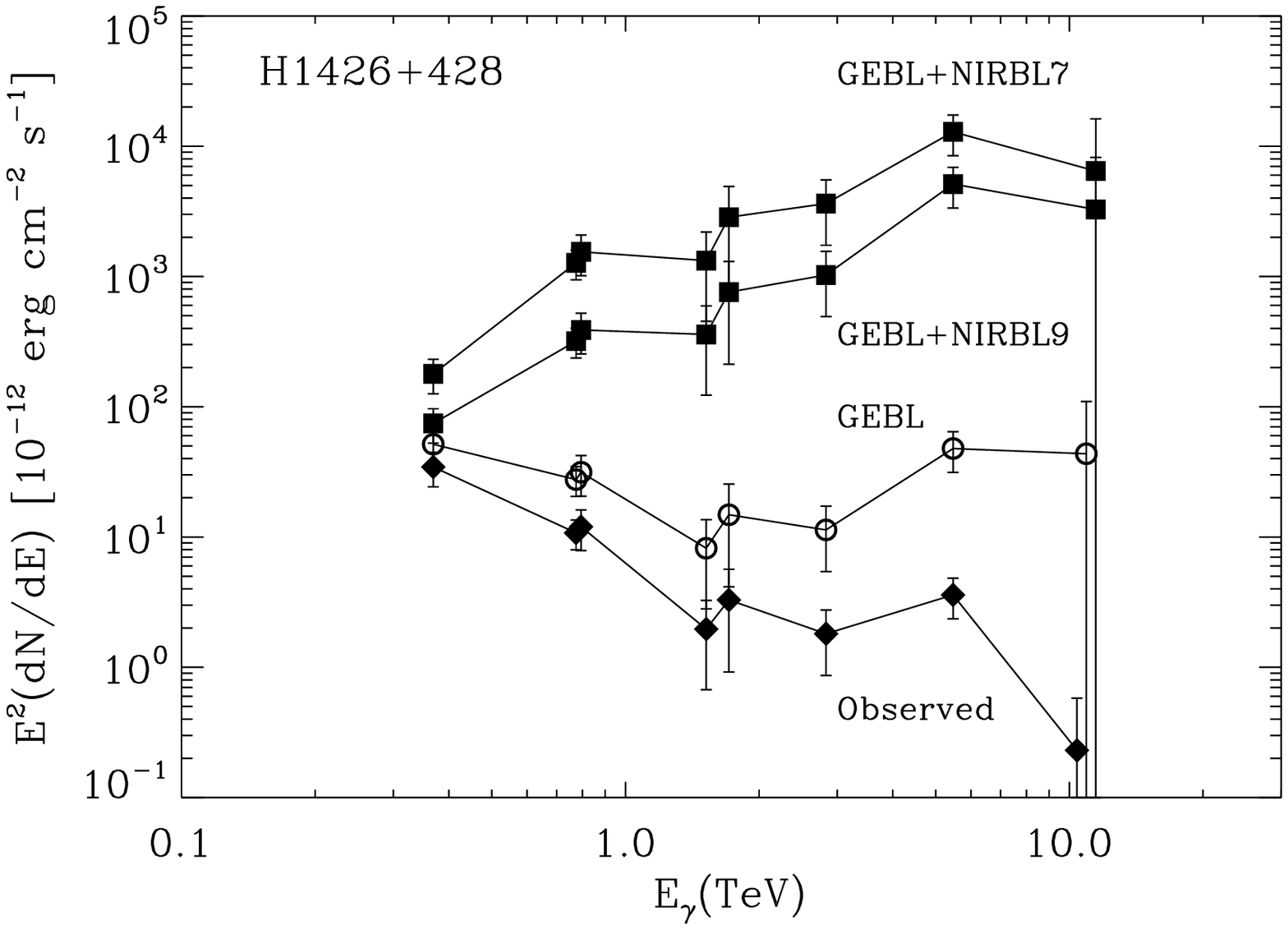}{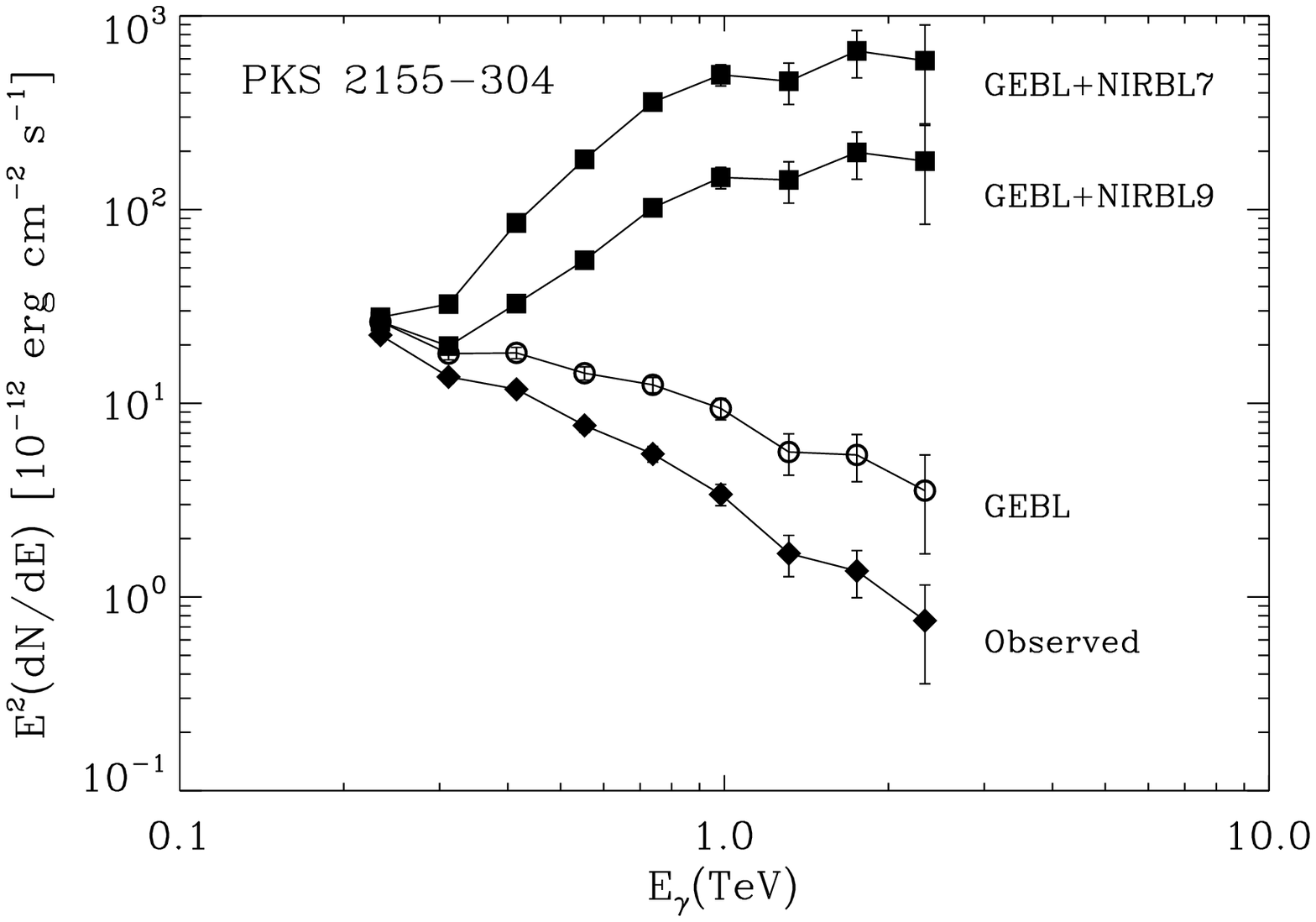}
 \caption{{\it Left panel}: The observed \gray\ spectrum of H1426+428 (Petry et al. 2002; filles diamonds) corrected for absorption by only the GEBL (open circles), the GEBL+NIRBL7 (filled squares), and the GEBL+NIRBL9 (open squares) realizations of the EBL. {\it Right panel}: The same as the left panel for PKS2155-304. The data point represent the H.E.S.S. observations by Aharonian et al. (2005)}
 \end{figure}

\subsection{PKS2155-304}

The situation is somewhat different for the blazar PKS2155-304, which was observed at $\sim$TeV energies by Chadwick et al. (1999), and more recently in 2002/2003 by the High Energy Stereoscopic System (HESS) collaboration (Aharonian et al. 2005). Compared to H1426+428, its observed \gray\ spectrum extends to lower energies (200~GeV), and is well fit by a power law to a higher degree of precision. So a priori, from knowledge of the effect of the NIRBL on the \gray\ opacity towards this blazar (Figures 3 and 4) we expect the NIRBL to create a break in its absorption-corrected \gray\ spectrum at around $\sim$ 300~GeV.  Indeed, Figure 6 (right panel) shows that if only corrected for GEBL absorption, the absorption-corrected blazar spectrum is still simply a power law, but when corrected for any of the GEBL+NIRBL realizations, the absorption-corrected blazar spectrum attains a parabolic form  with a peak at around 1-2~TeV, and an inflection at around 300~GeV.

Concentrating only on the $\sim$200 GeV to 3~TeV region of the spectrum, it seems difficult to ascertain the extragalactic nature of the NIRBL because, without a priori knowledge of the intrinsic spectrum of PKS2155-304, there is no reason for favoring any one of the absorption-corrected spectrum in Figure 6 over the other two. 
Additional $\lesssim$ 100~GeV observations are important for resolving this current ambiguity. For example, an observed 100~GeV flux continuing the observed 0.2 to 2 TeV power law trend would clearly eliminate the extragalactic interpretation of the NIRBL. The GEBL+NIRBL absorption-corrected spectrum of the blazar would then have a dip at $\sim$ 300~GeV, requiring an extraordinary coincidence for this EBL to have the exact spectral shape that compensates for this feature in the intrinsic blazar spectrum to produce an observed power law spectrum for this source.

  \begin{figure}
   \epsscale{0.75}
 \plotone{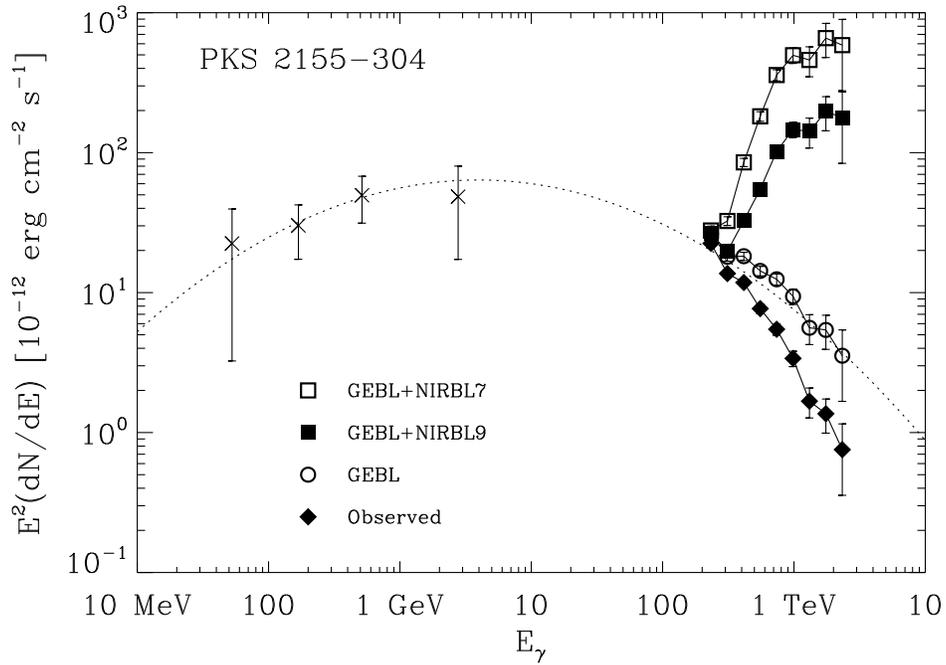}
  \caption{The observed \gray\ spectrum of PKS2155-304 (Aharonian et al. 2005) corrected for the various realizations of the EBL (see Figure 6) is compared to the EGRET data of Vestrand et al. (1995, X symbols). A detailed discussion of the figure is in the  text. }
   \epsscale{1.0}
 \end{figure}

The nature of the NIRBL becomes less ambiguous if data obtained by the EGRET experiment on  board the {\it Compton Gamma Ray Observatory} (Vestrand, Stacy, \& Sreekumar 1995) are included in the analysis. The entire non-thermal $\sim$10~eV to 3~TeV spectral energy distribution (SED) of PKS2155-304 can be well explained by the synchrotron self Compton (SSC) model. In this model, the very high energy (VHE) \gray s are produced by inverse Compton scattering of the synchrotron photons by the very same electrons that produce them. The SED is therefore characterized by a double peak: a synchrotron  peak located at UV-X-ray energies ($\sim$ 0.1 - 1~keV), and a Compton peak located at energies of about 0.1-1~TeV. SSC model fits to the SED of PKS2155-304 are presented in Figure 8 of Chiappetti et al. (1999). At the time their paper was written there was only one TeV datum point for this blazar (Chadwick et al. 1999).  Figure~7 shows the combined EGRET and HESS spectra of this blazar, which covers the most of the Compton region of the SED. The EGRET data are presented by "X", and the symbols for the different VHE \gray\ spectra are the same  as in Figure 6 (right panel). The UV to X-ray spectrum of the blazar is variable (see summary in Chiappetti et al. 1999), and since the EGRET and HESS data are not coeval, there may be a relative offset between the two spectra. 

The figure shows that if the observed PKS2155-304 spectrum is corrected for only GEBL absorption (open circles) then the resulting SED is characterized by a smooth parabolic function (dotted line) with a peak between $\sim$ 1 and 100~GeV, consistent with the SSC model presented by Chiappetti et al. (1999).
In contrast, corrected for the GEBL+NIRBL realization of the EBL, the resulting  0.1~GeV - 3~TeV SED shows a significant break between $\sim$ 50 and 200~GeV. The SED is initially relatively flat with a spectral index of $\alpha \approx 0.4$ between 0.1 and 10~GeV, followed by a very steep rise with $\sim E^2dN/dE\propto E^{2.3}$ at TeV energies (solid squares). 
As mentioned earlier, the MeV and TeV data are not contemporaneous, and it is possible that the SED has evolved in the time span between the observations so that  the SED from each individual epoch may be described by a simple SSC model. However, both, the MeV and TeV regions of the spectrum would have had to evolve in a very dramatic fashion, decreasing by over two orders of magnitude with a significant change in slope and peak energies, in order to create a smooth  function representing the inverse Compton (IC) part of the SED at both observing epochs.  
Using data obtained during several simultaneous multiwavelengths campaigns spanning the 1994 to 1999 time period, Ciprini \& Tosti (2003) fit the IC part of the SED with an SSC model that produced a Compton peak at about 40~GeV. 
If the 2002/2003 epoch is also described by a simple SSC model, then the GEBL+NIRBL absorption-corrected spectrum suggests that the inverse Compton peak shifted to $\sim$ 3~TeV. In an SSC model a shift of the IC peak by two orders of magnitude would also require a similar shift in synchrotron peak energy, since the X-ray band provides the seed photons that produce the TeV \gray s. Indeed, large correlated shifts in the synchrotron and IC peaks with flaring activity have been observed for Mrk~501 and 1ES~1959+650 (Costamante 2004). 
However, repeated observations over a $\sim$ 3 year time period have shown that the synchrotron peak of PKS2155-304 (as well as that of Mrk~421) is nearly stable at $\lesssim$ 1~keV.  (e.g. Ciprini \&
Tosti 2003;  Costamante 2003). PKS2155-304 may therefore have different physical characteristics than these two blazars, and the stability of its synchrotron  peak argues against the presence of strong variability in its IC peak as well. It seems therefore very unlikely that the GEBL+NIRBL absorption-corrected spectrum represents its intrinsic TeV spectrum.

Alternatively, the very steep rise in the TeV spectrum may be explained by ``hadronic" models. If the TeV \gray s are produced by hadronic interactions between or synchrotron emission from extremely high energy protons, it may be possible to reproduce a pileup of photons at TeV energies (e.g. Mannheim 1993, Aharonian 2000).  However, in these models, even a proton spectrum with a sharp pile-up produces a relatively flat intrinsic spectrum (see, for example, Figure 10 in Aharonian 2000).

\subsection{Mrk 421 and Mrk 501}

Figure 8 shows the observed and attenuation corrected spectrum of Mrk~421 and Mrk~501 for the different EBL realizations. The Mrk~501 observations were taken from Samuelson et al. (1998), and the Mrk~421 data were taken from Krennrich et al. (2001). Because of the  uncertainties in the observations, and the relative proximity of these sources, ($z \approx$ 0.03) the increased \gray\ opacity caused by the addition of the NIRBL to the GEBL  does not have a significant affect on the intrinsic blazar spectrum.  

  \begin{figure}
 \plottwo{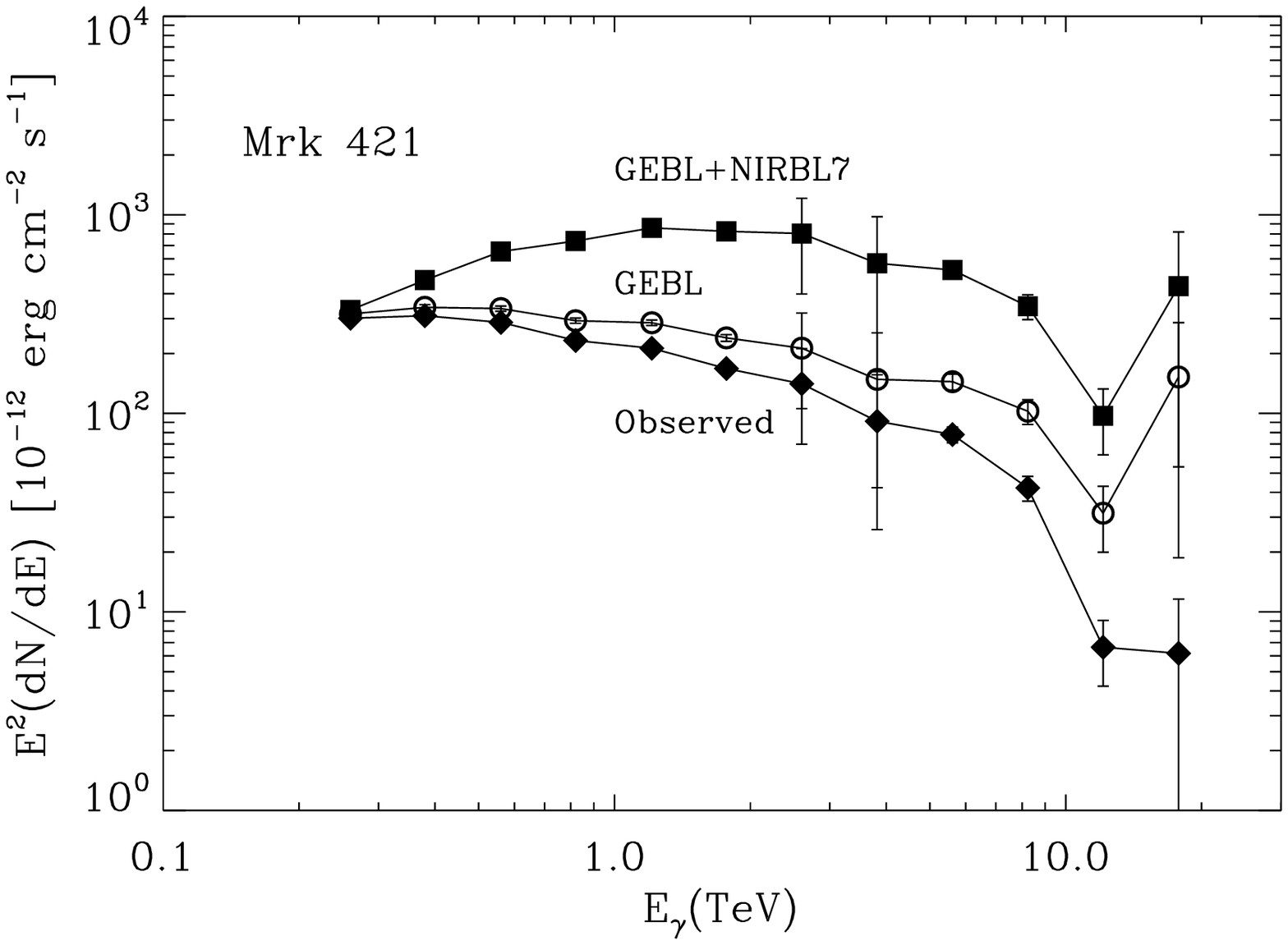}{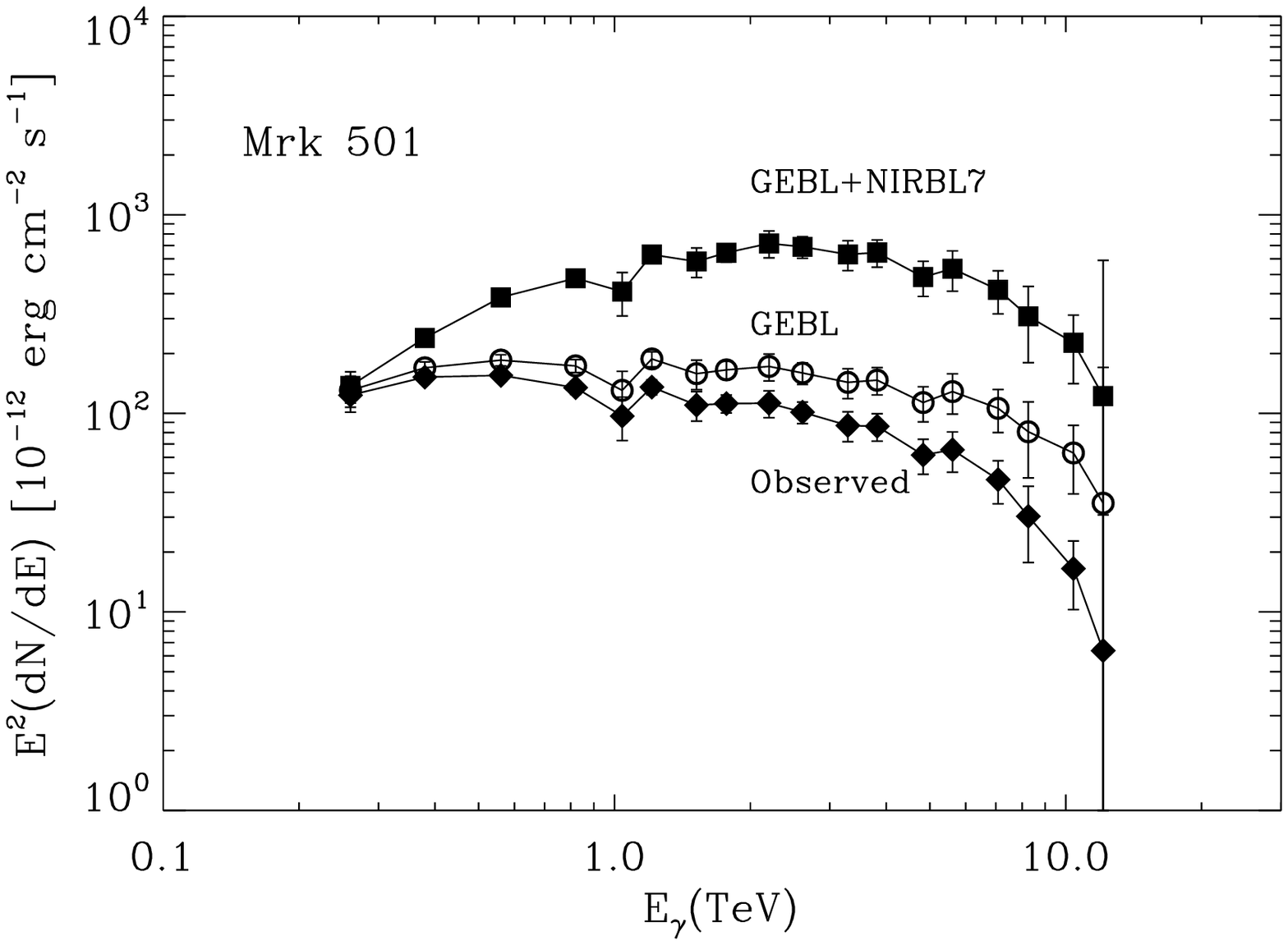}
  \caption{same as Figure 6, for the blazars Mrk~421 and Mrk~501. Because of their relative proximity, the effect of the EBL absorption has only a moderate effect on the observed blazar spectra.}
 \end{figure}

\section{SUMMARY}

The first generation of stars can have a significant effect on the spectral shape and intensity of the extragalactic background light (EBL), which may have been detected by the DIRBE and NIRS in the form of of an excess in the near-infrared background light (NIRBL) over that expected from normal star-forming galaxies (GEBL). If the NIRBL is indeed extragalactic, then it has a significant effect on the \gray\ opacity of blazars. The detectibility of the resulting absorption will depend on the distance of the blazar, and the quality of the \gray\ data.  

In general, the abrupt increase in the NIRBL intensity between 0.8 and 1.25 \mum\ (Figure 1) gives rise to a more gradual increase in the \gray\ opacity at around $\sim$ 300~GeV (Figures 3, 4). Given an observed blazar spectrum, this increased opacity will in turn create a break in the absorption-corrected blazar spectrum at energies around 300~GeV, which will not be present if the EBL consisted of only normal star forming galaxies.

An additional effect of the NIRBL is to increase the power required to be emitted by the source at $E \gtrsim$ 0.5~TeV in order to produce the observed spectrum in the presence of this EBL absorption feature. Such energy requirements can provide useful constraints on \gray\ production models.

In order to examine the possible signature of this unique spectral component on the TeV spectra of blazar, we calculated the absorption-corrected spectra of the blazars, H1426+428 ($z$ = 0.129), PKS2155-304 ($z$ = 0.117), Mrk~421 ($z$ = 0.030), and Mrk~501 ($z$ = 0.033), for several EBL realization, with and without the NIRBL. 

H1426+428, was observed over the 0.37 - 10.4~TeV energy range. Consequently, the attenuation corrected blazar spectrum did not exhibit any distinct features attributed to the NIRBL. In particular, the TeV spectrum of H1426+428 which Mapelli, Salvaterra, \& Ferrara (2004) attributed to NIRBL absorption producing an observed "S"-shaped spectrum, can also be produced in EBL models {\it without} the NIRBL component. Both the GEBL and the GEBL+NIRBL corrected spectra require the intrinsic spectrum to be a power law, the former characterized by a $E^2dN/dE \sim constant$, the latter by a rising $\propto E$ behavior (Figures 5 and 6 (left panel). In principle, both spectra could be the intrinsic blazar spectrum, however, the latter seems less likely considering the spectral characteristics of other blazars.

PKS2155-304 was observed between 0.23 and 2.3~TeV, a spectral region that covers (barely) the energy region in which the \gray\ opacity is affected by the NIRBL. Consequently, while the GEBL corrected flux can be characterized by a power law, the GEBL+NIRBL one is fairly flat between $\sim$ 200 -- 300~GeV and rising steeply as $E^{2.3}$ at higher energies (Figures 6, right panel). Since the observed spectrum of this blazar is very well represented by a power law, it will require a very fine tuning of its intrinsic spectrum to produce this power law with the additional TeV opacity generated by the NIRBL. 
Furthermore, even proton synchrotron models for the VHE \gray\ emission may have difficulties producing such a steep rise in the intrinsic blazar spectrum, favoring the GEBL absorption-corrected one as the intrinsic blazar spectrum.

A stronger argument against  the extragalactic nature of the NIRBL can be made on the basis of the combined MeV to TeV spectrum of the blazar (Figure 7), which covers most of the inverse Compton region of the spectrum in the synchrotron self Compton (SSC) \gray\ production model. Even though the MeV and TeV data are not contemporaneous, it will require a very  dramatic evolution in the relative intensity and spectral shape of these two spectral regions, and in the position of the Compton peak between the 1996/1997 and 2002/2003 epochs for the GEBL+NIRBL absorption-corrected spectrum to be consistent with the SSC model.  

Further $\sim$ 10 - 100~GeV and contemporaneous MeV observations are needed to provide a more detailed characterization of the effect of the NIRBL opacity on the intrinsic blazar spectra. Given several possible intrinsic blazar spectra, such as those presented in Figures 6 and 7, more detailed models will be required to confirm that any one of these spectra is physically more acceptable than others in order to prove or disprove the extragalactic nature of the NIRBL.
  
{\bf Acknowledgements}
We thank Werner Hofmann for his comments on the manuscript, and Dirk Petry and Wyston Benbow for providing us with the H1426+428 and PKS2155-304 spectra, respectively, in digital form. We thank the anonymous referee for his/her useful comments on the manuscript. ED acknowledges the support of NASA's LTSA 2004.

\clearpage


\begin{references}
\reference{ } Aharonian, F. A. 2000, New Astronomy, 5, 377
\reference{ } Aharonian, F. A., Akhperjanian, A. G., Barrio, J. A.,  Beilicke, M., Bernl\"or, K.  et al. 2002, \aap, 384, L23
\reference{ } Aharonian, F. A., Akhperjanian, A. G., Beilicke, M., Bernl'\"or, K., B\'orst, H.-G. et al. 2003, \aap, 403, 523
\reference{ } Aharonian, F. A., Akhperjanian, A. G., Aye, K.-M., Bazer-Bachi, A. R., A., Beilicke, M. et al. 2005, \aap, 430, 865
\reference{ } Arendt, R. G., \& Dwek, E. 2003, \apj, 585, 305
\reference{ } Bennett, C. L. et al. 2003, \apj, 148, 1 
\reference{ } Bernstein, R. A., Freedman, W. L., \&  Madore, B. F. 2002, \apj,
571, 56
\reference{ } Cambr\'esy, L., Reach, W. T., Beichman, C. A., \& Jarrett, T. H.
2001, \apj, 555, 563
\reference{ } Chiappetti, L., Maraschi, L., Tavecchio, F., Celotti, A., Fosati, G. et al. 1999, \apj, 521, 552
\reference{ } Ciprini, S., \& Tosti, G. 2003, in "High Energy Blazar Astronomy", eds. A. Sillanpaa, L. O. Takalo, \& E. Valtaoja, ASP Conf. Ser. Vol. 299, 269 (astro-ph/0303107) 
\reference{ } Costamante, L. 2004, New Astron. Rev. 48, 498
\reference{ } Costamante, L., Aharonian, F., Horns, D., \& Ghisellini, G. 2004, New Astron. Rev., 48, 469
\reference{ } Dwek, E., \& Arendt, R. G. 1998, \apj, 508, L9
\reference{ } Dwek, E., \& Krennrich, F. 2005, \apj, 618, 657
\reference{ } Dwek, E., \& Arendt, R. G. 2005, in preparation
\reference{ } Fazio, G., Ashby, M. L. N., Barmby, P., Hora, J. L., Huang, J.-S. et al. 2004, \apjs, 154, 39
\reference{ } Gorjian, V., \& Wright, E. L., \& Chary, R. R. 2000, \apj, 536, 550
\reference{ } Hauser, M.G. et al. 1998, \apj, 508, 25
\reference{ } Krennrich, F., Badran, H. M., Bond, I. H., Bradbury, S. M., Buckley, J. H. et al. 2001, \apj, 560, L45
\reference{ } Hauser, M.G. \& Dwek, E. 2001, ARA\&A, 39, 249
\reference{ } Lagache, G., Haffner, L. M., Reynolds, R. J., \& Tufte, S. L.
2000, \aap, 354, 247
\reference{ } Mannheim K. 1993, \aap, 269, 67
\reference{ } Mapelli, M., Salvaterra, R., \& Ferrara, A. 2004, \mnras, submitted (astro-ph/0410615)
\reference{ } Matsumoto, T., Matsuura, S., Murakami, H., Tanaka, M., Fruend, M. et al. 2004, astro-ph/0411593
\reference{ } Metcalfe, L., Kneib, J.-P, McBreen, B., Altieri, B., Biviano, A. et al. 2003, \aap, 407, 791
\reference{ } Papovich, C., Dole, H., Egami, E., LeFloc'h, E., P\'erez-Gonz\'alez, P. G. et al. 2004, ApJS, 154, 70
\reference{ } Petry, D., Bond, I. H., Bradbury, S. M., Buckley, J. H., Carter-Lewis, D. A. et al. 2002, \apj, 580, 104
\reference{ } Santos, M. R., Bromm, V., \& Kamionkowski, M. 2002, \mnras, 336, 1082
\reference{ } Salvaterra, R., \& Ferrara, A. 2003, \mnras, 339, 973
\reference{ } Samuelson, F. W. et al. 1998 \apj, 501, L17 
\reference{ } Tavecchio,  F., Maraschi, L., \& Ghisellini, G. 1998, \apj, 509, 608
\reference{ } Totani, T., Yoshii, Y., Iwamuro, F., Maihara, T., \& Motohara, K. 2001, \apj, 550, L137
\reference{ } Vestrand, W. T., Stacy, J. G., \& Sreekumar, P. 1995, \apj, 454, L93
\reference{ } Wright, E. L., \& Reese, E. D. 2000, \apj, 545, 43
\reference{ } Wright, E. L. 2001, \apj, 553, 538
\end{references}
\end{document}